\documentclass{JHEP3}
\pdfoutput=1

\usepackage{graphicx}
\usepackage{amsmath,amsfonts,bbm,times,slashed,dcolumn}

\DeclareMathOperator{\re}{Re}
\DeclareMathOperator{\tr}{Tr}

\DeclareMathOperator{\diag}{\textrm{diag}}

\DeclareMathOperator{\Disc}{Disc}

\newcommand{\h}{\mathcal H}
\newcommand{\y}{\Upsilon}

\newcommand{\Zz}{\mathbbm{Z}}

\newcommand{\abs}[1]{{\vert #1 \vert}}

\renewcommand\epsilon\varepsilon
\renewcommand\phi\varphi
\renewcommand{\O}{\mathcal{O}}

\title{The epsilon expansion at next-to-next-to-leading order with
  small imaginary chemical potential}
\author{Christoph Lehner$\,^{a,b}$, Shoji Hashimoto$\,^c$, and Tilo Wettig$\,^a$\\
  \llap{$^a$}Institute for Theoretical Physics, University of Regensburg, 93040 Regensburg, Germany\\
  \llap{$^b$}RIKEN/BNL Research Center, Brookhaven National Laboratory, Upton, NY-11973, USA\\
  \llap{$^c$}High Energy Accelerator Research Organization (KEK), Tsukuba 305-080, Japan\\
  Email: \email{clehner@quark.phy.bnl.gov},
  \email{shoji.hashimoto@kek.jp},
  \email{tilo.wettig@physik.uni-regensburg.de}}

\abstract{We discuss chiral perturbation theory for two and three
  quark flavors in the epsilon expansion at next-to-next-to-leading
  order (NNLO) including a small imaginary chemical potential.  We
  calculate finite-volume corrections to the low-energy constants
  $\Sigma$ and $F$ and determine the non-universal modifications of
  the theory, i.e., modifications that cannot be mapped to random
  matrix theory (RMT).  In the special case of two quark flavors in an
  asymmetric box we discuss how to minimize the finite-volume
  corrections and non-universal modifications by an optimal choice of
  the lattice geometry.  Furthermore we provide a detailed calculation
  of a special version of the massless sunset diagram at finite
  volume.}


\preprint{April 30, 2010, KEK-CP-234, RBRC-842}

\begin{document}

\section{Introduction}
At low energies the theory of quantum chromodynamics (QCD) can be
described by a chiral effective theory.  If the theory is confined to
a finite volume and considered for small quark masses, the
$\epsilon$-regime power counting applies and replaces the standard
$p$-regime power counting, see Ref.~\cite{Gasser:1987ah}.  The
corresponding systematic expansion is called $\epsilon$-expansion.

At leading order (LO) in the $\epsilon$-expansion the theory becomes
zero-dimensional and is therefore described by chiral RMT
\cite{Shuryak:1992pi}, see \cite{Verbaarschot:2000dy,Akemann:2007rf}
for reviews.  The dimensionless quantities of RMT are mapped to the
dimensionful quantities of the chiral effective theory using the LO
low-energy constants (LECs) $\Sigma$ and $F$, see, e.g.,
Ref.~\cite{Basile:2007ki}.  In this way $\Sigma$ and $F$, which are of
great phenomenological importance, can be obtained from lattice QCD
simulations in the $\varepsilon$-regime by a fit to RMT predictions.
While $\Sigma$ can be determined rather easily, e.g., from the
distribution of the small Dirac eigenvalues, the extraction of $F$ is
more complicated and can be done, e.g., by the inclusion of a suitable
chemical potential \cite{Damgaard:2005ys,Akemann:2006ru} or by using
twisted boundary conditions \cite{Mehen:2005fw}.

Lattice QCD simulations in the $\varepsilon$-regime using an exactly
chiral lattice fermion formulation are feasible, and recently such a
set of lattice configurations, which also includes the effect of
dynamical quarks, was generated by the JLQCD and TWQCD collaborations
\cite{Fukaya:2007fb,Fukaya:2007yv,Fukaya:2009fh}.  In an upcoming
publication \cite{Lehner:2010yy} we further investigate the eigenvalue
spectrum of the Dirac operator on these configurations.

For typical volumes of state-of-the-art lattice QCD simulations,
finite-volume corrections to the RMT predictions cannot be neglected.
These corrections can be calculated in higher orders of the
$\varepsilon$-expansion.  At next-to-leading order (NLO) the
analytical results of RMT still apply, but the mapping of RMT to chiral
perturbation theory is modified, i.e., the LECs $\Sigma$ and $F$ are
replaced by finite-volume effective LECs $\Sigma_\text{eff}$ and
$F_\text{eff}$, see
Refs.~\cite{Damgaard:2007xg,Akemann:2008vp,Lehner:2009pz}.  At NNLO,
chiral perturbation theory can no longer be mapped to RMT, i.e.,
non-universal modifications appear.  These non-universal modifications
determine the systematic errors in fits of lattice data to RMT
predictions.

In this paper we calculate the finite-volume corrections at NNLO in
the $\varepsilon$-expansion in Euclidean space-time.  We allow for
nonzero imaginary chemical potential and consider its contribution to
leading order.  In this way we extend results of
Hansen~\cite{Hansen:1990un}, which were obtained without chemical
potential.  The paper is structured as follows.  In Sec.~\ref{sec:eps}
we calculate the finite-volume effective action and the corresponding
finite-volume effective low-energy and high-energy constants (HECs) at
NNLO.  In Sec.~\ref{sec:optnf2} we discuss how to minimize the
systematic deviations from RMT as well as the finite-volume
corrections to $\Sigma$ and $F$ in the special case of two
quark flavors on an asymmetric lattice by an optimal choice of the
lattice geometry.  We conclude
in Sec.~\ref{sec:conc} and provide a detailed calculation of a special
version of the massless sunset diagram at finite volume in
App.~\ref{app:p6}.

\section{The finite-volume effective theory at NNLO}\label{sec:eps}
In this section we discuss the finite-volume effective theory at NNLO
in the $\epsilon$-expansion.  To this end we give the bare Lagrangian
of the $p$-expansion at NLO and the LO contribution of the invariant
integral measure in Sec.~\ref{sec:barelagr}.  In Sec.~\ref{sec:fvact}
we define the finite-volume effective action, and in
Sec.~\ref{sec:fvlec} we calculate the finite-volume effective LECs and
HECs. The LECs $L_1,\ldots,L_8$ and the HEC $H_2$ that appear in the bare
Lagrangian of the $p$-expansion at NLO are scale-dependent.  We
renormalize the theory in Sec.~\ref{sec:fvlec} and confirm that the
scale dependence of $L_1,\ldots,L_8$ and $H_2$ is the
same as in the ordinary $p$-expansion \cite{Gasser:1984gg}.

\subsection{The bare Lagrangian and the path-integral measure}\label{sec:barelagr}
We parametrize the Nambu-Goldstone manifold of chiral
perturbation theory with $N_f$ quark flavors by
\begin{align}\label{eqn:defu}
  U(x) = U_0 \exp\biggl[\frac{i\sqrt 2}{F} \xi(x) \biggr]\,,
\end{align}
where $\xi$ is a complex matrix in flavor space of dimension $N_f$
with $\xi=\xi^\dagger$ and $\tr \xi = 0$.  The constant mode is
separated in $U_0$, and thus $\int d^4x\, \xi(x) = 0$, see
Ref.~\cite{Gasser:1987ah}.  For nonzero imaginary chemical potential
the LO Lagrangian of the effective theory in Euclidean space-time is given by
\begin{align}
  {\cal L}_1 = \frac{F^2}4 \tr[ \nabla_\rho U(x)^{-1}
  \nabla_\rho U(x) ] - \frac{\Sigma}{2} \tr[ M^\dagger
  U(x) + U(x)^{-1} M]
\end{align}
with
\begin{align}\label{eqn:nablachem}
  \nabla_\rho U(x) = \partial_\rho U(x) -i \delta_{\rho 0}[C,U(x)]\,,
\end{align}
where $M=\diag(m_1,\ldots,m_{N_f})$, $m_f$ is the mass of quark flavor
$f$, $C=\diag(\mu_1,\ldots,\mu_{N_f})$, and $i\mu_f$ is the imaginary
chemical potential of quark flavor $f$, see, e.g.,
Ref.~\cite{Akemann:2008vp}.  We consider very small quark masses such
that the Compton wavelength of the pions, given by the
Gell-Mann--Oakes--Renner relation
\begin{align}
  m_\pi^2 = \frac{2m_f\Sigma}{F^2}\,,
\end{align}
exceeds the size of the space-time box.  This defines the
$\varepsilon$-regime of QCD that was first discussed in
Ref.~\cite{Gasser:1987ah}. 
The corresponding $\varepsilon$-regime power counting
\cite{Gasser:1987ah} is defined by
\begin{align}\label{eqn:defepspc}
  V \sim \varepsilon^{-4}\,,\qquad M \sim \varepsilon^4\,,\qquad C
  \sim \varepsilon^2\,,\qquad
  \partial_\rho \sim \varepsilon\,,\qquad \xi(x) \sim \varepsilon
\end{align}
which gives a consistent perturbative expansion in $\varepsilon^2 \sim
1/F^2\sqrt{V}$, see, e.g., Ref.~\cite{Lehner:2009pz}.  The
$\varepsilon$-expansion is applicable if the quantities
\begin{align}\label{eqn:defepsilon}
  m_f V \Sigma\quad\text{and}\quad \mu_f^2 F^2 V
\end{align}
are not larger than $\mathcal O(1)$ and the volume is
sufficiently large, i.e.,
\begin{align}
  4\pi F^2\sqrt{V} \gg 1\,,
\end{align}
see also the quantitative discussion in Sec.~\ref{sec:optnf2}.

In order to include all NNLO contributions in the $\epsilon$-expansion
we need to include the NLO Lagrangian of the $p$-expansion which for
$N_f=3$ is given by
\begin{align}\label{eqn:ltwo}
  {\cal L}_2 &= -L_1 [\tr (\nabla_\mu U^{-1}\nabla^\mu U)]^2
  -L_2 \tr (\nabla_\mu U^{-1}\nabla_\nu U) \tr(\nabla^\mu U^{-1}\nabla^\nu U) \notag\\
  &\quad-L_3 \tr (\nabla_\mu U^{-1}\nabla^\mu U \nabla_\nu
  U^{-1}\nabla^\nu U) \notag\\&\quad +\left(\frac{2\Sigma}{F^2}\right)
  L_4 \tr (\nabla_\mu U^{-1}\nabla^\mu U) \tr(M U^{-1} + M^\dagger
  U)\notag\\&\quad +\left(\frac{2\Sigma}{F^2}\right) L_5 \tr
  [\nabla_\mu U^{-1}\nabla^\mu U(M U^{-1} + M^\dagger
  U)] \notag\\&\quad -\left(\frac{2\Sigma}{F^2}\right)^2 L_6 [\tr
  (M U^{-1} + M^\dagger U)]^2 
  -\left(\frac{2\Sigma}{F^2}\right)^2 L_7 [\tr (M U^{-1} -
  M^\dagger U)]^2\notag\\&\quad
  -\left(\frac{2\Sigma}{F^2}\right)^2 L_8\tr (M U^{-1}M U^{-1} +
  M^\dagger U M^\dagger U)
  -\left(\frac{2\Sigma}{F^2}\right)^2 H_2 \tr (M^\dagger M)\,,
\end{align}
where $H_2$ is a high-energy constant (HEC) corresponding to a contact
term that is needed in the renormalization of one-loop graphs, see
Refs.~\cite{Hansen:1990un} and \cite{Gasser:1984gg}.  The
field-strength tensors are not included since they vanish in the case
of a constant vector source \cite{Gasser:1984gg}, of which an
imaginary chemical potential is a special case, and therefore the LECs
$L_9$ and $L_{10}$ and the HEC $H_1$ do not appear.  Note that in the
case of $N_f=2$ not all of the terms in ${\cal L}_2$ are independent
and the LECs $L_1,\ldots,L_{10}$ can be mapped to the LECs
$l_1,\ldots,l_7$ as described in Ref.~\cite{Bijnens:1999hw}.  In the
case of $N_f>3$ one needs to include additional terms in
Eq.~\eqref{eqn:ltwo}, see, e.g., Ref.~\cite{Bijnens:1999hw}.

The invariant measure relevant to NNLO in the $\epsilon$-expansion is
given by
\begin{align}
  d[U] &= d[U_0] d[\xi] \left( 1 - \frac{N_f}{3F^2 V} \int d^4x
    \tr[\xi(x)^2] \right)\,,
\end{align}
see Ref.~\cite{Hansen:1990un}, where $d[U_0]$ is the invariant measure
of the constant mode and $d[\xi]$ is the flat measure of the fields
$\xi$.

\subsection{The finite-volume effective action}\label{sec:fvact}
The volume dependence of the theory is contained entirely in the
propagators of the fields $\xi$ \cite{Hasenfratz:1989pk}, and
therefore we can obtain a finite-volume effective action in terms of
the constant mode $U_0$ by averaging over the fluctuations in $\xi$.
At NLO this leads to a finite-volume effective action that differs
from the LO (RMT) result by finite-volume corrections to the LECs
$\Sigma$ and $F$, see
Refs.~\cite{Damgaard:2007xg,Akemann:2008vp,Lehner:2009pz}.  We now perform
the expansion of the partition function
\begin{align}
  Z=\int d[U]\,e^{-\int d^4x\,(\mathcal L_1+\mathcal L_2)}
\end{align}
in terms of fields $\xi$ to NNLO and average over the
fields using computer algebra,\footnote{We use a C++ library for
  tensor algebra developed by one of the authors.}  resulting in
\begin{align}
  Z=\int d[U_0]\,e^{-S_\text{eff}}\,.
\end{align}
The finite-volume effective action $S_\text{eff}$ contains invariant
terms including the constant mode $U_0$, the mass matrix $M$, and the
chemical potential matrix $C$.  At NNLO in the $\varepsilon$-expansion
and to leading order $C^2$ in the imaginary chemical potential,
$S_\text{eff}$ can be written as
\begin{align}
  S_\text{eff} &= -\frac{V\Sigma_\text{eff}} 2\tr( M^\dagger U_0
  \notag + U_0^{-1} M )-\frac{VF^2_\text{eff}}2 \tr( C
  U_0^{-1} C U_0 )\notag\\&\quad
 +\y_1
  \Sigma (VF)^2   \tr(C)[\tr(U_0\{M^\dagger, C\})+\tr(U_0^{-1}\{C,M\})]
  \notag\\&\quad
  +\y_2 \Sigma (VF)^2\tr(\{M^\dagger, C\}U_0 C + \{C, M\} C U_0^{-1}
  \notag\\&\qquad\qquad\qquad\qquad + \{U_0, C\} U_0^{-1} C U_0
  M^\dagger + CU_0 \{C,U_0^{-1}\} M U_0^{-1} ) \notag\\&\quad
 +\y_3
  \Sigma (VF)^2 \tr(U_0^{-1}CU_0C + C^2) \tr(M U_0^{-1} +
  M^\dagger U_0) \notag\\&\quad
 +\y_4 \Sigma (VF)^2 \tr(U_0^{-1}CU_0C - C^2) \tr(M
  U_0^{-1} + M^\dagger U_0)\notag \\&\quad +\y_5 \Sigma (VF)^2\tr(
  [M^\dagger, C] U_0 C + [C,M]C U_0^{-1}
  \notag\\&\qquad\qquad\qquad\qquad + [U_0, C] U_0^{-1}C U_0 M^\dagger
  + CU_0 [C,U_0^{-1}] M U_0^{-1} ) \notag\\&\quad+\y_6
  (V\Sigma)^2 [\tr(M U_0^{-1}+M^\dagger U_0)]^2 \notag\\&\quad +\y_7
  (V\Sigma)^2 [\tr(M U_0^{-1} - M^\dagger U_0)]^2 \notag\\&\quad +\y_8
  (V\Sigma)^2 [\tr(M U_0^{-1}MU_0^{-1})+\tr(M^\dagger U_0M^\dagger
  U_0)] \notag\\&\quad +\h_1VF^2\tr( C^2 )
  +\h_2 (V\Sigma)^2 \tr(M^\dagger M) + \h_3 V F^2 (\tr C)^2\,,
\end{align}
where $\{A,B\} = A B + B A$ and $\Sigma_\text{eff}$, $F_\text{eff}$,
and $\y_1,\ldots,\y_8$ are finite-volume effective LECs that will be
given in Sec.~\ref{sec:fvlec}.  Note that the terms corresponding to
$\h_1$, $\h_2$, and $\h_3$ do not couple to $U_0$, and therefore the
$\h_i$ can be viewed as finite-volume effective HECs.  The terms
proportional to the $\h_i$ only contain external sources and are
therefore not relevant for low-energy phenomenology, but they are
needed for the computation of operator expectation values, see
Ref.~\cite{Bijnens:2006zp}.  They are also relevant to the
renormalization of the coupling constants discussed in
Sec.~\ref{sec:fvlec}.  The finite-volume effective HECs $\h_i$ are
also given in Sec.~\ref{sec:fvlec}.

Unlike the LECs $L_1,\ldots,L_8$ and the HEC $H_2$, the finite-volume
effective LECs $\y_1,\ldots,\y_8$ and HECs $\h_1$, $\h_2$, and $\h_3$ are
finite and depend on the volume.  Specifically, we show in
Sec.~\ref{sec:fvlec} that
\begin{align}
  \Sigma_\text{eff},F_\text{eff},\h_1 = {\cal O}(\epsilon^0)\,,\qquad
  \h_3 = {\cal O}(\varepsilon^2)\,,\qquad
  \h_2,\y_1\,,\ldots\,,\y_8 = {\cal O}(\varepsilon^4)\,,
\end{align}
i.e., $\h_2$, $\h_3$, and $\y_1,\ldots,\y_8$ vanish in the
infinite-volume limit.

The terms corresponding to the $\y_i$ cannot be mapped to RMT, see,
e.g., Ref.~\cite{Basile:2007ki}.  Therefore their magnitude quantifies
the systematic deviations from RMT at finite volume.  We will return
to this point at the end of Sec.~\ref{sec:optnf2}.

\subsection{The finite-volume effective low-energy and high-energy constants}\label{sec:fvlec}
In the following we express the finite-volume effective LECs
$\Sigma_\text{eff}$, $F_\text{eff}$, and $\y_1$, $\ldots$, $\y_8$ and
the finite-volume effective HECs $\h_1$, $\h_2$, and $\h_3$ in terms
of shape coefficients $P_1,\ldots,P_6$ defined below.  The resulting
expressions for the effective LECs and HECs are given in terms of the
massless finite-volume propagator in dimensional regularization,
\begin{align}
  \bar G(x) = \frac1V \sum_{k \ne 0} \frac{e^{ikx}}{k^2}\,,
\end{align}
where the sum is over all nonzero momenta, see
Ref.~\cite{Hasenfratz:1989pk}.  We use the identity
\begin{align}
  \partial_\rho^2 \bar G(x) \bigr\vert_{x=0} = \frac1V
\end{align}
and finally express the result in terms of the shape coefficients
\begin{align}\label{eqn:defallprop}
  P_1 &= V \partial_0^2\bar G(0)\,, &  P_2 &= \sqrt V \bar G(0)\,,\notag\\
  P_3 &= \sqrt V \int d^dx [\partial_0^2\bar G(x)] \bar G(x)\,, &
  P_4 &=  \int d^dx \,\bar G(x)^2\,,\notag\\
  P_5 &=  \int d^dx \,d^dy [\partial_0^2\bar G(x+y)] \bar G(x) \bar G(y)\,,\  &
  P_6 &= V \int d^dx [\partial_0^2\bar G(x)] \bar G(x)^2\,,
\end{align}
where $d$ is the number of space-time dimensions.  We use
conservation of momentum, by which all one-loop propagators can be related to
\cite{Hasenfratz:1989pk}
\begin{align}
\label{eqn:grb}
  \bar G_r &= \frac{\Gamma(r)}{V} \sum_{k \ne 0} \frac1{(k^2)^r}\,,
\end{align}
where $r \in \mathbbm{R}$ and $\Gamma(r)$ is the Gamma function
\cite{abramowitz+stegun}.  The partial derivatives can be evaluated
using
\begin{align}
  \label{eqn:partial}
  \partial_{\tilde{L}_\nu}(\tilde{L}_\nu \bar G_r) &= \frac{2\Gamma(r+1)}{V} \sum_{k \ne
    0} \frac{k_\nu^2}{(k^2)^{r+1}}\,,
\end{align}
where $\tilde{L}_\nu$ is the length of the space-time box in dimension
$\nu=0,1,2,3$ and $\partial_{\tilde{L}_\nu}$ denotes the partial
derivative w.r.t.~$\tilde{L}_\nu$.  Note that in Eq.~\eqref{eqn:partial} no sum over $\nu$
is implied.  The shape coefficients $P_1,\ldots,P_5$ contain only
one-loop propagators and are given by
\begin{align}
  P_1 &= -\frac V2 \partial_{\tilde{L}_0} (\tilde{L}_0 \bar G_0)\,, &  
  P_2 &= \sqrt V \bar G_1\,, \notag\\
  P_3 &= -\frac{\sqrt{V}}2 \partial_{\tilde{L}_0} (\tilde{L}_0 \bar G_1)\,, &
  P_4 &= \bar G_2\,, \notag\\
  P_5 &= -\frac14\partial_{\tilde{L}_0} (\tilde{L}_0 \bar G_2)\,.
\end{align}
For convenience we state the result of Ref.~\cite{Hasenfratz:1989pk}
explicitly as
\begin{align}\label{eqn:defapppre}
  \bar G_r &= \lim_{m \to 0}\left[ \frac1{(4\pi)^{d/2}} \Gamma(r-d/2) (m^2)^{d/2-r} + g_r - \frac{\Gamma(r)}{V m^{2r}} \right], \notag\\
  V g_0 &= \beta_0 + \beta_1 m^2 \sqrt V + \frac12 \beta_2 m^4 V - \log(m^2\sqrt V) \notag\\&\quad
  + \frac{V m^4}{2(4\pi)^2}\left[\log(m^2\sqrt V) - \frac12\right] + \O(m^6)\,,\notag\\
  g_{r+1} &= -\frac{\partial g_r}{\partial(m^2)}
\end{align}
with shape coefficients $\beta_n$ given in Eq.~(B.14) of Ref.~\cite{Hasenfratz:1989pk}.  We express $\bar G_0$, $\bar G_1$,
and $\bar G_2$ in terms of $\beta_0$, $\beta_1$, and $\beta_2$ and
find
\begin{align}\label{eqn:propresa}
  P_1 &= -\frac 12 \tilde{L}_0\partial_{\tilde{L}_0}\beta_0 +\frac14\,, &  P_2  &= -\beta_1\,,\notag\\
  P_3 &= \frac14 \beta_1 +\frac12 \tilde{L}_0 \partial_{\tilde{L}_0} \beta_1\,, & P_4 &= -2\lambda + \beta_2 + \frac{\log\sqrt{V}}{(4\pi)^2}\,, \notag\\
  P_5 &= -\frac14 P_4 -\frac14 \tilde{L}_0\partial_{\tilde{L}_0}\beta_2 -
  \frac2{(16\pi)^2}\,,
\end{align}
where we borrow the definition of $\lambda$ from
Ref.~\cite{Gasser:1984gg},
\begin{align}\label{eqn:poleresa}
  \lambda &= \frac{\mu^{d-4}}{(4 \pi)^2}
  \left\{\frac1{d-4}-\frac12\left[1+\Gamma'(1)+\log(4 \pi)\right]\right\}\notag\\
  &\to \frac1{(4 \pi)^2}
  \left\{\frac1{d-4}-\frac12\left[1+\Gamma'(1)+\log(\mu^2)+\log(4 \pi)\right]\right\}\,.
\end{align}
We explicitly include the dependence on the scale $\mu$, which we
define with mass dimension one.  Note that the logarithms of
dimensionful quantities in Eqs.~\eqref{eqn:propresa} and
\eqref{eqn:poleresa} can always be combined to logarithms of
dimensionless quantities.  The shape coefficient $P_6$ is
calculated in App.~\ref{app:p6}, see Eq.~\eqref{eq:p6}.  It contains a special version of the
massless sunset diagram at finite volume. 

In the following we state the resulting expressions for
$\Sigma_\text{eff}$, $F_\text{eff}$, $\y_1,\ldots,\y_8$, $\h_1$,
$\h_2$, and $\h_3$.  The finite-volume effective chiral condensate is
given by
\begin{align}\label{eqn:nnlosigma}
 \frac{\Sigma_\text{eff}}{\Sigma} &= 1 - \frac{P_2}{F^2\sqrt{V}} (N_f-N_f^{-1})
- \frac12(1 - N_f^{-2}) \frac{P_2^2}{F^4 V} \notag\\&\quad
+ \frac{P_4}{F^4 V}( N_f^2 - 1 )
+\frac8{F^4 V}\bigl[ (N_f^2-1) L_4 + (N_f-N_f^{-1}) L_5 \bigr]\,,
\end{align}
which agrees with Eqs.~(22) and (23) of Ref.~\cite{Hansen:1990un}.
The finite-volume effective LEC
\begin{align}\label{eqn:nnlof}
 \frac{F_\text{eff}^2}{F^2}  &= 1 - 2 N_f \frac{P_2}{F^2\sqrt{V}} - 2 N_f \frac{P_3}{F^2\sqrt{V}}
  + 2 N_f^2 \frac{P_2 P_3}{F^4 V} + 2 N_f^2 \frac{P_3^2}{F^4 V} + N_f^2 \frac{P_2^2}{F^4 V}   \notag\\&\quad
  + N_f^2 \frac{(2P_4 + 4P_5 + P_6)}{F^4 V}+ \frac{16}{F^4 V} \bigl[ (N_f^2-1) L_1 + L_2 +
    (N_f-N_f^{-1}) L_3\bigr] \notag\\&\quad+ \frac{16 P_1}{F^4 V} \bigl[ 2 L_1 +
    N_f^2 L_2 + (N_f - 2N_f^{-1})L_3\bigr]
\end{align}
and the finite-volume effective HEC
\begin{align}\label{eqn:nnloftilde}
  \h_1 &= \frac12 + N_f \frac{P_3}{F^2 \sqrt{V}} - N_f^2 \frac{P_2 P_3}{F^4 V}
  - N_f^2 \frac{P_3^2}{F^4 V} + \frac12 N_f^2 \frac{(P_6 - 4P_5)}{F^4 V}
  \notag\\&\quad + \frac{8}{F^4 V} \bigl[ (N_f^2-1) L_1 + L_2 +
    (N_f-N_f^{-1}) L_3\bigr] \notag\\& \quad+ \frac{8 P_1}{F^4 V} \bigl[ 2
    L_1 + N_f^2 L_2 + (N_f - 2N_f^{-1})L_3\bigr]
\end{align}
contain the contribution of the two-loop propagator in $P_6$.  The
other finite-volume effective LECs and HECs are given by
\begin{align}\label{eqn:nnloupsilon1}
  \y_1 &= \frac12 \frac{P_4+4 P_5}{F^4 V} \,, &
  \y_2 &=  -\frac14 N_f \y_1\,, \notag\\
  \y_3 &= -\frac12 \y_1\,, &
  \y_4 &= -\frac14 \frac{P_4}{F^4 V} -\frac{4L_4}{F^4 V}\,,\notag\\
  \y_5 &= -\frac18 N_f \frac{P_4}{F^4 V} - \frac{2L_5}{F^4 V}\,, &
  \y_6 &= -\frac18\bigl(1+2 N_f^{-2}\bigr) \frac{P_4}{F^4 V} -
  \frac{4L_6}{F^4 V} \,,\notag \\
  \y_7 &= -\frac{4 L_7}{F^4 V}\,, & \y_8 &= \frac12 \bigl(N_f^{-1} -
  \frac14 N_f\bigr) \frac{P_4}{F^4 V} - \frac{4L_8}{F^4 V}
\end{align}
and
\begin{align}\label{eqn:nnloupsilon2}
  \h_2 &= \bigl(N_f^{-1} - \frac14 N_f\bigr) \frac{P_4}{F^4 V}
  -\frac{4H_2}{F^4 V} \,, \notag\\
  \h_3 &= -\frac{(P_2+2P_3)}{F^2 \sqrt{V}} + \frac{N_f}2
  \frac{(P_2^2+4 P_3^2)}{F^4 V} + N_f \frac{(P_4+4P_5)}{F^4 V}
  + 2 N_f \frac{P_2 P_3}{F^4 V}\,.
\end{align}

Note that the shape
coefficients $P_4,P_5$, and $P_6$ as well as $H_2$ and the $L_i$
are divergent and need to be
renormalized.  We separate their scale dependence as
\begin{align}
  P_4 &= P_4^r - 2 \lambda\,,\qquad
  P_5 = P_5^r + \frac12 \lambda\,,\qquad
  P_6 = P_6^r + \frac13 \lambda - \frac{10}3 P_1 \lambda\,,\notag\\
  L_i & = L_i^r + \Gamma_i \lambda\;\;\;\text{with}\;\;\;i=1,\ldots,8\,,\qquad
  H_2 = H_2^r + \Delta_2 \lambda\,,
\end{align}
where the quantities with superscript $r$ are finite.  For $N_f=3$ the
divergences in Eqs.~\eqref{eqn:nnlosigma} - \eqref{eqn:nnloupsilon2}
can be absorbed if and only if we choose
\begin{align}
  \Gamma_4 &= \frac18\,, &
  \Gamma_5 &= \frac38\,, &
  \Gamma_6 &= \frac{11}{144} \,, \notag\\
  \Gamma_7 &= 0 \,, &
  \Gamma_8 &= \frac5{48}\,, &
  \Delta_2 &= \frac5{24}\,,
\end{align}
and
\begin{align}\label{eqn:renormg13}
  \frac{30}{16} = 16 \Gamma_1+2\Gamma_2+\frac{16}3 \Gamma_3 = 2
  \Gamma_1 + 9 \Gamma_2 + \frac73 \Gamma_3\,.
\end{align}
The coefficients $\Gamma_4,\ldots,\Gamma_8$ and $\Delta_2$ are equal
to the coefficients obtained in the one-loop expansion in the $p$
power counting, see Ref.~\cite{Gasser:1984gg}.  The renormalization
conditions of Eq.~\eqref{eqn:renormg13} for
$\Gamma_1,\Gamma_2,\Gamma_3$ are also compatible with the result of
Ref.~\cite{Gasser:1984gg},
\begin{align}
  \Gamma_1 = \frac3{32}\,,\qquad \Gamma_2 = \frac3{16}\,,\qquad
  \Gamma_3 = 0\,.
\end{align}
Note that the divergences in $\h_3$, $\y_1$, $\y_2$, and
$\y_3$ cancel independently of the choice of $\Gamma_i$.

To summarize, we obtain finite expressions for the finite-volume
effective LECs and HECs in Eqs.~\eqref{eqn:nnlosigma} -
\eqref{eqn:nnloupsilon2} if we replace $P_4,P_5$, $P_6$, $H_2$ and the
$L_i$ by their corresponding renormalized parts with superscript $r$.
Note that the dependence on the scale $\mu$ drops out of the final
results for the finite-volume effective LECs and HECs.  The
renormalization in the case of $N_f=2$ is discussed in
Sec.~\ref{sec:optnf2}.

\section{Optimal geometries: Two quark flavors in an asymmetric box}\label{sec:optnf2}
In the following we discuss the finite-volume corrections to $\Sigma$
and $F$ and the coefficients of the non-universal terms at NNLO in the
$\varepsilon$-expansion.  We explicitly consider the case of $N_f=2$
and an asymmetric box with lattice geometries
\begin{align}
  (a_x)\qquad \tilde{L}_0 &= x L\,,\qquad \tilde{L}_1 = \tilde{L}_2 = \tilde{L}_3 = L\,, \notag\\
  (b_x)\qquad \tilde{L}_3 &= x L\,,\qquad \tilde{L}_0 = \tilde{L}_1 =
  \tilde{L}_2 = L\,,
\end{align}
where $x \in \{1, 3/2, 2, 3, 4\}$.  The three-flavor coupling
constants $L_1,\ldots,L_5$ can be related to the two-flavor coupling
constants $l_1$, $l_2$, and $l_4$ by
\begin{align}
  l_1 &= 4 L_1 + 2 L_3\,, & l_2 &= 4 L_2 \,, & l_4 &=8 L_4 + 4 L_5\,,
\end{align}
see, e.g., Eqs.~(3.15) and (3.16) of Ref.~\cite{Bijnens:1999hw}.
Therefore
\begin{align}\label{eqn:sigmaefftwoflavor}
  \frac{\Sigma_\text{eff}}{\Sigma} &= 1 -
  \frac{3P_2}{2F^2\sqrt{V}} - \frac{3P_2^2}{8F^4 V} +
  \frac{3P_4}{F^4 V} +\frac{3 l_4}{F^4 V}
\end{align}
and
\begin{align}\label{eqn:fefftwoflavor}
  \frac{F_\text{eff}^2}{F^2} &= 1 - \frac{4P_2}{F^2\sqrt{V}} -
  \frac{4P_3}{F^2\sqrt{V}} + \frac{8 P_2 P_3+8 P_3^2+4 P_2^2}{F^4 V} +
  \frac{8P_4 + 16P_5 + 4 P_6}{F^4 V} \notag\\& \quad+ \frac{12 l_1 + 4
    l_2}{F^4 V} + \frac{P_1( 8 l_1 + 16 l_2)}{F^4 V}\,.
\end{align}
In Ref.~\cite{Gasser:1983yg} the scale dependence of the coupling
constants $l_i$ with $i=1,\ldots,7$ is separated as
\begin{align}
  l_i = l_i^r + \gamma_i \lambda\,,
\end{align}
where
\begin{align}
  \gamma_1=\frac 13\,,\qquad \gamma_2=\frac 23\,, \qquad \gamma_4 = 2\,.
\end{align}
It is straightforward to check that the divergences in
Eqs.~\eqref{eqn:sigmaefftwoflavor} and \eqref{eqn:fefftwoflavor}
cancel with this set of $\gamma_1$, $\gamma_2$, and $\gamma_4$.  
We therefore obtain finite results for $\Sigma_\text{eff}$ and
$F_\text{eff}$ if we replace $P_4,P_5,P_6$ and $l_1,l_2,l_4$ in
Eqs.~\eqref{eqn:sigmaefftwoflavor} and \eqref{eqn:fefftwoflavor} by
their corresponding renormalized parts with superscript $r$.

The shape coefficients $P_1,P_2,P_3$ and the renormalized shape
coefficients $P_4^r,P_5^r,P_6^r$ at scale $\mu=V^{-1/4}$ are given in
Table~\ref{tab:sc} for geometries $(a_x)$ and $(b_x)$.  The details of
the calculation of $P_6^r$ are given in App.~\ref{app:p6}.
\newcolumntype{d}{D{.}{.}{2.6}}
\newcolumntype{e}{D{.}{.}{2.8}}
\newcolumntype{h}{D{.}{.}{2}}
\newcolumntype{g}{D{.}{.}{3}}
\TABULAR[ht]{|c|ddddde|}{\hline & \multicolumn1h{P_1} & \multicolumn1h{P_2} & \multicolumn1h{P_3} & \multicolumn1h{P_4^r} & \multicolumn1h{P_5^r} &
  \multicolumn1{g|}{P_6^r} \\\hline
  $(a_1)$ & 0.249999 & -0.140461 & 0.035115 & -0.020305 & 0.004285 & -0.0106(1) \\
  $(a_{3/2})$ & -0.251470 & -0.123339 & -0.014414 & -0.018457 & 0.001121 & 0.00940(4) \\
  $(a_2)$ & -0.674807 & -0.083601 & -0.076021 & -0.012954 & -0.005093 & 0.02487(2) \\
  $(a_3)$ & -1.512610 & 0.041942 & -0.237477 & 0.012150 & -0.031163 & 0.0319(1)\\
  $(a_4)$ & -2.350148 & 0.215097 & -0.440882 & 0.062628 & -0.082324 & -0.0214(3)\\
  $(b_{3/2})$ & 0.417156 & -0.123339 & 0.045917 & -0.018457 & 0.004723 & -0.01513(3)\\
  $(b_2)$ & 0.558268 & -0.083601 & 0.053207 & -0.012954 & 0.004960 & -0.01577(7)\\
  $(b_3)$ & 0.837535 & 0.041942 & 0.065178 & 0.012150 & 0.005282 &  -0.0080(1)\\
  $(b_4)$ & 1.116713 & 0.215097 & 0.075261 & 0.062628 & 0.005510 &
  0.0117(1)\\\hline}{\label{tab:sc}Shape coefficients $P_1,P_2,P_3$ and
  renormalized shape coefficients $P_4^r,P_5^r,P_6^r$ at scale
  $\mu=V^{-1/4}$ for geometries $(a_x)$ and $(b_x)$.  The error in the
last column is due to the extrapolation described in
App.~\ref{app:complete}.} 

The renormalized coupling constants $l_i^r$ can be related to
scale-independent constants $\bar l_i$ by
\begin{align}
  l_i^r = \frac{\gamma_i}{2(4\pi)^2}\bigl[ \bar l_i + \log(m_\pi^2 / \mu^2) \bigr],
\end{align}
where $m_\pi$ is the mass of the pion, see Ref.~\cite{Gasser:1983yg}, and
\begin{align}
  \bar l_1 = -0.4 \pm 0.6\,,\qquad \bar l_2 = 4.3 \pm 0.1\,, \qquad
  \bar l_4 = 4.4 \pm 0.2\,,
\end{align}
see Ref.~\cite{Bijnens:2006zp}.  Therefore
\begin{align}
  l_i^r &= \frac{\gamma_i}{2(4\pi)^2}\bigl[ \bar l_i + \log(m_\pi^2
    \sqrt{V}) \bigr]
\end{align}
at scale $\mu=V^{-1/4}$.  Note again that the finite-volume corrections to
$\Sigma$ and $F$ are independent of the choice of scale $\mu$.

\TABULAR[t]{|c|ddddd|}{\hline & \multicolumn1h{(a_1)} & \multicolumn1h{(a_{3/2})} & \multicolumn1h{(a_2)} &
  \multicolumn1h{(a_3)} & \multicolumn1{h|}{(a_4)} \\\hline
  $\Sigma^\text{NLO}_\text{eff}/\Sigma$ & 1.3455 & 1.2477 & 1.1454 & 0.9404 & 0.7355\\
  $\Sigma^\text{NNLO}_\text{eff}/\Sigma$ & 1.39(1) & 1.288(7) & 1.202(5) & 1.047(3)& 0.906(3) \\
  $F^\text{NLO}_\text{eff}/F$ & 1.3004 & 1.3182 & 1.3192 & 1.3193 & 1.3193\\
  $F^\text{NNLO}_\text{eff}/F$ & 1.279(9) & 1.305(4) & 1.306(2) &
  1.292(1) & 1.261(2) \\\hline & & \multicolumn1h{(b_{3/2})} & \multicolumn1h{(b_2)} &
  \multicolumn1h{(b_3)} & \multicolumn1{h|}{(b_4)} \\\hline
  $F^\text{NLO}_\text{eff}/F$ & & 1.1894 & 1.06816 & 0.7710 & 0.2186 \\
  $F^\text{NNLO}_\text{eff}/F$ & & 1.182(8) & 1.092(7) & 0.959(6) &
  0.919(5) \\\hline }{\label{tab:effres} Finite-volume corrections to
  $\Sigma$ and $F$ at NLO and at NNLO for geometries $(a_x)$ and
  $(b_x)$ at $m_\pi^2\sqrt{V} = 1$, $F=90$ MeV, and $L=1.71$ fm.  The
  error in $\Sigma^\text{NNLO}_\text{eff}$ is due to the uncertainty
  in $\bar l_4$, the error in $F^\text{NNLO}_\text{eff}$ is due to the
  uncertainty in $\bar l_1$, $\bar l_2$, and $P_6^r$.}
In Table~\ref{tab:effres} we give explicit values for the finite-volume
corrections to $\Sigma$ and $F$ at NLO and NNLO for geometries $(a_x)$
and $(b_x)$ with $m_\pi^2\sqrt{V} = 1$, $F=90$ MeV, and $L=1.71$ fm,
which roughly corresponds to the values of the JLQCD lattice simulations.
Note that the $\varepsilon$-expansion converges well for this set of
parameters as long as the asymmetry of the lattice is not too strong
(convergence is worst in geometry $(b_4)$).  We calculate the
NLO result by discarding terms of order $1/F^4V$ in
Eqs.~\eqref{eqn:sigmaefftwoflavor} and \eqref{eqn:fefftwoflavor}.
Note that for the same value of $x$,
$\Sigma_\text{eff}$ is independent of the choice of lattice geometry
$(a_x)$ or $(b_x)$.  
\FIGURE[t]{ \centering
  \begin{tabular}{ccc}
  \includegraphics{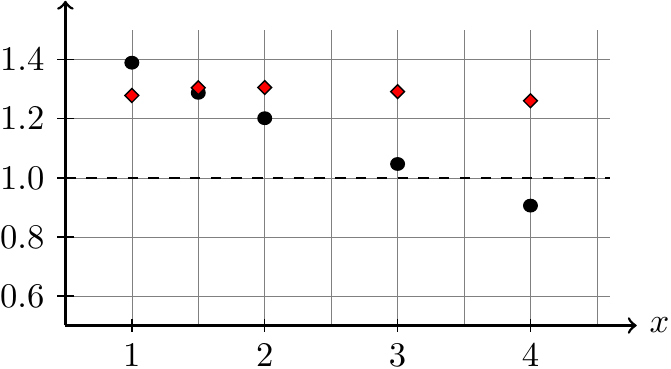} & &  \includegraphics{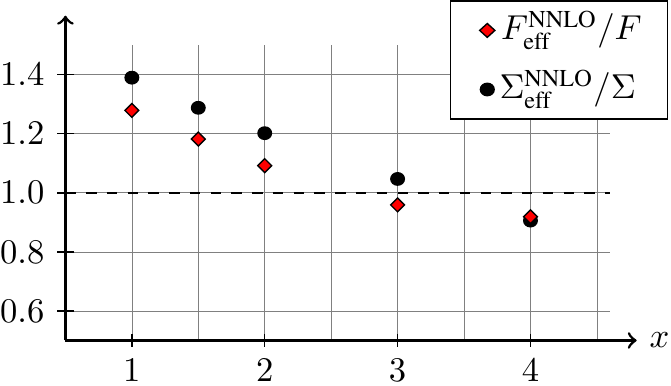}\\
  $(a_x)$ & &$(b_x)$
  \end{tabular}
  \caption{Finite-volume corrections to $\Sigma$ and $F$ at NNLO for
    geometries $(a_x)$ and $(b_x)$ at $m_\pi^2\sqrt{V} = 1$, $F=90$
    MeV, and $L=1.71$ fm.}
  \label{fig:effnnlo}
}
In Figure~\ref{fig:effnnlo} we visualize the NNLO results of
Table~\ref{tab:effres}.  We confirm the picture obtained in
Ref.~\cite{Lehner:2009pz} at NLO that the finite-volume corrections to
$F$ can be largely reduced by an asymmetric lattice geometry with one
appropriately large spatial dimension instead of one large temporal dimension.

We now turn to the non-universal terms that cannot be mapped to RMT.
It follows from Eqs.~\eqref{eqn:propresa} and \eqref{eqn:nnloupsilon1}
that the coefficients $\y_4,\ldots,\y_8$ are independent of the choice
of lattice geometry $(a_x)$ or $(b_x)$ for the same value of $x$.  The
coefficients $\y_1$, $\y_2$, and $\y_3$, however, are affected by the
choice of lattice geometry $(a_x)$ or $(b_x)$, and we have
\begin{align}
 \y_1, \y_2, \y_3 \propto  P^r_4 + 4 P^r_5\,.
\end{align}
We give values for $P^r_4 + 4 P^r_5$ in Figure~\ref{fig:nonun} at
scale $\mu=V^{-1/4}$ for lattice geometries $(a_x)$ and $(b_x)$.
\FIGURE[!b]{ \centering
  \includegraphics{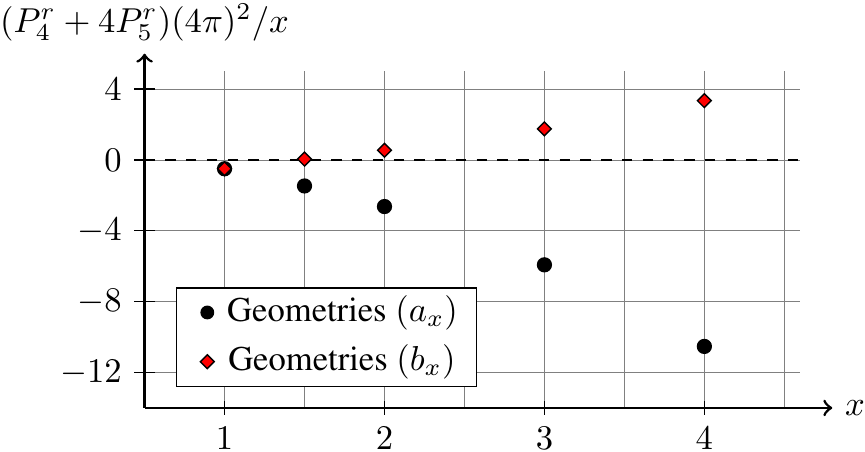}
  \caption{Linear combination of shape coefficients $P^r_4 + 4 P^r_5$
    for different geometries at scale $\mu=V^{-1/4}$.  We divide by
    $x$ since $V=xL^4$ in Eq.~\eqref{eqn:nnloupsilon1}.}
  \label{fig:nonun}
} Note that the non-universal contribution of $\y_1$, $\y_2$, and
$\y_3$ is reduced significantly in lattice geometry $(b_x)$ compared
to lattice geometry $(a_x)$ for the same value of $x$.  In an upcoming
publication \cite{Lehner:2010yy} we perform the corresponding lattice
simulation for $x=2$ and show numerically that the systematic
deviations from RMT are indeed smaller for lattice geometry $(b_2)$
compared to lattice geometry $(a_2)$.

\section{Conclusions}\label{sec:conc}
We discussed the $\epsilon$-expansion at NNLO and determined the
finite-volume effective action and the finite-volume effective LECs
and HECs to this order.  In the special case of two dynamical quarks
confined to an asymmetric box we have confirmed the picture obtained
at NLO that finite-volume corrections to the LECs $\Sigma$ and $F$ can
be significantly reduced by choosing one appropriately large spatial
dimension instead of a large temporal dimension, see
Ref.~\cite{Lehner:2009pz}.  Furthermore, we have shown that the
systematic deviations from random matrix theory can also be reduced in
the setup with one large spatial dimension.  This implies that in
order to determine the LECs $\Sigma$ and $F$ from eigenvalue
correlation functions, as suggested in
Refs.~\cite{Damgaard:2005ys,Akemann:2006ru} and performed in a pilot
study in Ref.~\cite{DeGrand:2007tm}, one should choose an asymmetric
lattice with one large spatial dimension.  This will be demonstrated
numerically in an upcoming publication \cite{Lehner:2010yy}.

We would like to add that even though we did not explicitly perform
our calculations in a partially quenched setup, it is straightforward
to extend our results to the partially quenched case, see, e.g.,
Ref.~\cite{Lehner:2009pz} for a discussion at NLO, and we expect our
findings to be unmodified by the presence of valence quarks.

\acknowledgments 

We thank Hidenori Fukaya for stimulating discussions.  Two of us (CL
and TW) are grateful to the Theory Group of the IPNS, KEK for their
hospitality.  This work was supported in part by BayEFG (CL), the
Grant-in-Aid (No. 21674002) of the Japanese Ministry of Education
(SH), and DFG and KEK (TW).

\appendix
\section{The massless sunset diagram at finite volume}\label{app:p6}
In this section we calculate the two-loop contribution defined by
\begin{align}
  P_6 &= V \int d^dx [\partial_0^2\bar G(x)] \bar G(x)^2 = -\frac1V \sum_{k \ne
    0} \frac{k_0^2}{k^2} \sum_{p\ne 0, -k} \frac1{p^2(p+k)^2}\,,
\end{align}
where the sum is over all nonzero momenta $k$ and $p$ with $p+k\ne0$, and $k_0$ is
the temporal component of the momentum vector $k$.  We first express
the propagators without constant mode as the limit of ordinary,
massive propagators,
\begin{align}
  P_6 = \lim_{m \to 0} P_6(m^2) = \lim_{m \to 0} \left[ P_6^0(m^2) + P_6^1(m^2) \right]
\end{align}
with
\begin{align}
   P_6^0(m^2) &=\frac2{m^2V} \sum_{k} \frac{k_0^2}{(k^2+m^2)^2}\,, \notag\\
   P_6^1(m^2) &=-\frac1V \sum_{k,p} \frac{k_0^2}{(p^2+m^2)[(p+k)^2+m^2](k^2+m^2)}\,.
\end{align}
The terms $P_6^0(m^2)$ and $P_6^1(m^2)$ are calculated separately in
the following.

\subsection{The term $P_6^0(m^2)$}
We partition the term $P_6^0(m^2)$ into its infinite-volume part and the
finite-volume propagator $g_1$ defined in Eq.~\eqref{eqn:defapppre}.
We find
\begin{align}
  P_6^0(m^2) &=\frac2{m^2V} \sum_{k} \frac{k_0^2}{(k^2+m^2)^2}
  =\frac1{m^2}\partial_{\tilde{L}_0}\biggl(\tilde{L}_0\frac1V \sum_{k} \frac1{k^2+m^2}\biggr) \notag\\
  &=\frac1{m^2} \frac1{(4\pi)^{d/2}} \Gamma(1-d/2) (m^2)^{d/2-1}
   +\frac1{m^2} g_1(m^2) +\frac1{m^2}\tilde{L}_0\partial_{\tilde{L}_0}g_1(m^2)\,,
\end{align}
where
\begin{align}
  g_1(m^2) &= \frac1{Vm^2} -\frac{\beta_1}{\sqrt V} -\frac{m^2 \log
    (m^2 \sqrt V)}{(4\pi)^2} -m^2 \beta_2 + {\cal O}(m^4)\,.
\end{align}
Therefore we can express $P_6^0(m^2)$ in terms of $\beta_1$, $\beta_2$, and
$\lambda$ as
\begin{align}
  P_6^0(m^2) &=2 \lambda -\frac1{m^2\sqrt V}\tilde{L}_0\partial_{\tilde{L}_0}\beta_1
  -\frac{\beta_1}{2m^2 \sqrt V} -\frac{\log \sqrt V}{(4\pi)^2}
-\partial_{\tilde{L}_0}(\tilde{L}_0\beta_2) -\frac1{2(4\pi)^2} +
  {\cal O}(m^2)\,.
\end{align}

\subsection{The term $P_6^1(m^2)$}
The second term
\begin{align}
  P_6^1(m^2) &= -\frac1V \sum_{k} \frac{k_0^2}{k^2+m^2}
    \sum_{p} \frac1{(p^2+m^2)[(p+k)^2+m^2]}
\end{align}
is more involved.  We use Poisson's summation formula
\begin{align}\label{eqn:poissonfrm}
  \sum_{n=-\infty}^\infty e^{2 \pi i n \phi} = \sum_{n=-\infty}^\infty \delta(\phi - n)
\end{align}
and write
\begin{align}
  P_6^1(m^2) = -V \sum_{r,s} \int
  \frac{d^dk}{(2\pi)^d}&\frac{d^dp}{(2\pi)^d} \exp\biggl[i \sum_j \tilde{L}_j
    (r_j k_j+ s_j p_j)\biggr] \notag\\&\times
  \frac{k_0^2}{(p^2+m^2)[(p+k)^2+m^2](k^2+m^2)}\,,
\end{align}
where the sum is over $r,s \in \Zz^4$.  We partition the sum over $r$
and $s$ into
\begin{align}
 (A)  & \qquad r = 0 \land s = 0\,, \notag\\
 (B) &\qquad  r \ne 0 \land s = 0 \,, \notag\\
 (C) & \qquad r = 0 \land s \ne 0\,, \notag\\
 (D) &\qquad r \ne 0 \land s \ne 0 \land s = r\,, \notag\\
 (E) &\qquad r \ne 0 \land s \ne 0 \land s \ne r \,.
\end{align}
Part (A) is given by the infinite-volume sunset diagram, see
Ref.~\cite{Amoros:1999dp}, which scales with $V m^d$ and therefore
vanishes in the massless limit.  The parts $P_6^{1B}(m^2),\ldots,P_6^{1E}(m^2)$
are calculated in the following.

\subsubsection{The term $P_6^{1B}(m^2)$}
Along the lines of Eqs.~(A.10) and (A.11) of
Ref.~\cite{Colangelo:2006mp} we separate
\begin{align}
  P_6^{1B}(m^2) &= P_6^{1B1}(m^2) + P_6^{1B2}(m^2)
\end{align}
with
\begin{align}
  P_6^{1B1}(m^2) &= -V \sum_{r \ne 0} \int
  \frac{d^dk}{(2\pi)^d}\frac{d^dp}{(2\pi)^d}
  \frac{k_0^2}{k^2+m^2}\frac1{(p^2+m^2)^2}\exp\biggl(i
    \sum_j \tilde{L}_j r_j k_j\biggr),\notag\\
  P_6^{1B2}(m^2) &= -V \sum_{r \ne 0} \int
  \frac{d^dk}{(2\pi)^d}\frac{d^dp}{(2\pi)^d}
  \frac{k_0^2}{k^2+m^2}\exp\biggl(i \sum_j \tilde{L}_j r_j k_j\biggr)
  \notag\\&\quad\times\biggl[
  \frac1{(p^2+m^2)[(p+k)^2+m^2]}-\frac1{(p^2+m^2)^2}\biggr]\,.
\end{align}
The term $P_6^{1B1}(m^2)$ contains the ultraviolet divergence and can be
calculated explicitly,
\begin{align}
  P_6^{1B1}(m^2) &= -V \int\frac{d^dp}{(2\pi)^d}\frac1{(p^2+m^2)^2} \sum_{r
    \ne 0} \int \frac{d^dk}{(2\pi)^d} \frac{k_0^2}{k^2+m^2}\exp\biggl(i
    \sum_j \tilde{L}_j r_j k_j\biggr) \notag\\
  &=-2 \lambda P_1 -\frac{1+\log m^2}{(4\pi)^2}P_1 +{\cal O}(V
  m^d)\,,
\end{align}
where $P_1$ is the one-loop shape coefficient defined in
Eq.~\eqref{eqn:defallprop}.  The term $P_6^{1B2}(m^2)$ is finite.
After a tedious but straightforward calculation along the lines of
Ref.~\cite{Hasenfratz:1989pk} we can express $P_6^{1B2}(m^2)$ as
\begin{align}\label{eq:tedious}
 P_6^{1B2}(m^2) =-\frac{1}{(8\pi)^2}\sum_{r \ne 0}
\int_0^\infty  dx dy dz \: K(x,y,z) \: \exp\biggl[-(x+y+z) \frac{m^2 \sqrt{V}}{4 \pi}
\biggr]
\end{align}
with
\begin{align}
  K(x,y,z) &= 
\frac{1}{(x y+x z+y z)^{3}}
\biggl[2 (x + y)
 - \frac{({\tilde{L}_0}^2/\sqrt{V}) [2 r_0 (x + y)]^2\pi}{(y z + x y + x z)}
\biggr] \notag\\&\quad\times
\exp\biggl[
 - \sum_j\frac{({\tilde{L}_j}^2/\sqrt{V}) r_j^2 (x + y)\pi}{(y z + x y + x z)}\biggr]\notag\\&\quad
-\frac1{(x+y)^{2}z^{3}}
\Bigl[2-({\tilde{L}_0}^2/\sqrt{V}) 4\pi r_0^2/z\Bigr]
\exp\biggl[
-\sum_j ({\tilde{L}_j}^2/\sqrt{V}) \pi \frac{r_j^2}{z}\biggr]\,.
\end{align}
This expression is suitable for a numerical evaluation of $P_6^{1B2}(m^2)$
if we perform the integral over $x$, $y$ and $z$ in spherical
coordinates.  In Sec.~\ref{app:p61e} we discuss how to efficiently
calculate infinite sums such as the sums over $r_0,\ldots,r_3$ with
$r\ne 0$ in Eq.~\eqref{eq:tedious}.

\subsubsection{The term $P_6^{1C}(m^2)$}
The method used to separate the divergent part of $P_6^{1B}(m^2)$ does not
work for the integral over $k$ since it has a power divergence.
Nevertheless, we can calculate the divergent sub-diagram
\begin{align}
  I_{\mu\nu}(m,p) &=  \int \frac{d^dk}{(2\pi)^d} \frac{k_\mu k_\nu}{(k^2+m^2)[(p+k)^2+m^2]}
\end{align}
explicitly.  The result is given by \cite{Passarino:1978jh}
\begin{align}
  I_{\mu\nu}(m,p) &=g_{\mu\nu}\int_0^1 dx \frac{[m^2+x(1-x)
    p^2]\log[m^2+x(1-x) p^2]}{2(4\pi)^2} \notag\\&\quad
  -p_\mu p_\nu \int_0^1 dx \frac{x^2}{(4\pi)^2} \bigl[1+\log[m^2+x(1-x)p^2]\bigr] \notag\\
  &\quad +g_{\mu\nu}\lambda \biggl(\frac16 p^2+m^2\biggr) -\frac23
  \lambda p_\mu p_\nu\,.
\end{align}
We can thus separate the divergent part of
\begin{align}
  P_6^{1C}(m^2) &=-V \sum_{s \ne 0} \int \frac{d^dp}{(2\pi)^{d}}
  \frac{I_{00}(m,p)}{p^2+m^2}\exp\biggl(i \sum_j \tilde{L}_j s_j
      p_j\biggr)\,,
\end{align}
which is given by
\begin{align}
  [P_6^{1C}(m^2)]_\text{UV} &=-\frac{\lambda V}6 \sum_{s \ne 0} \int
  \frac{d^dp}{(2\pi)^{d}} \frac{ (p^2+m^2) +5 m^2 -4
    p_0^2}{p^2+m^2}\exp\biggl(i \sum_j \tilde{L}_j s_j
    p_j\biggr) \notag \\
  &= -\frac56\lambda - \frac23 \lambda P_1 + {\cal O}(m^2)\,.
\end{align}
In the calculation of $[P_6^{1C}(m^2)]_\text{UV}$ we used the identities
\begin{align}
  V \sum_{s \ne 0} \int \frac{d^dp}{(2\pi)^{d}}
  \frac1{p^2+m^2}\exp\biggl(i \sum_j \tilde{L}_j s_j p_j\biggr) &= \frac1{m^2} + {\cal O}(m^0)\,,\notag\\
  \sum_{s \ne 0}\int \frac{d^dp}{(2\pi)^{d}} \:\exp\biggl(i \sum_j \tilde{L}_j s_j p_j\biggr) &= 0\,,\notag\\
  \sum_{s \ne 0}\int \frac{d^dp}{(2\pi)^{d}}
  \frac{p_0^2}{p^2+m^2}\exp\biggl(i \sum_j \tilde{L}_j s_j p_j\biggr) &= -P_1
  + {\cal O}(m^d)\,.
\end{align}
Note that the first two identities hold for arbitrary $d$.  Thus there
is no finite contribution from the product of these integrals with
$\lambda$.

The finite contributions to $P_6^{1C}(m^2)$ are given by
\begin{align}
  [P_6^{1C}(m^2)]_\text{finite} &=-\frac{V}{2(4\pi)^2} \sum_{s \ne 0} J'_s(m^2)\,,
\end{align}
where
\begin{align}
 J'_s(m^2) =\int_0^1 dx \int \frac{d^4p}{(2\pi)^4} \frac{{\cal F}_1(m^2,p^2)-p_0^2{\cal F}_2(m^2,p^2)}
{p^2+m^2}\exp\biggl(i \sum_j \tilde{L}_j s_j p_j\biggr)
\end{align}
with
\begin{align}
 {\cal F}_1(m^2,p^2) &= [m^2+x(1-x)p^2]\log[m^2+x(1-x) p^2]\,,\notag\\
 {\cal F}_2(m^2,p^2) &= 2 x^2 \big(1+\log[m^2+x(1-x)p^2]\big)\,.
\end{align}
We define $L^s_i = \tilde{L}_i s_i$ and rotate the coordinate system of $p$
such that
\begin{align}
  J'_s(m^2) &= \int_0^1 dx \int \frac{d^4p}{(2\pi)^4}
  \frac{{\cal F}_1(m^2,p^2)+(1/ {\tilde{L}_0}^2){\cal F}_2(m^2,p^2)\partial_{s_0}^2 }{p^2+m^2}
  \exp(i L^s p_0)
\end{align}
with $L^s = [\sum_{n=0}^3(L^s_n)^2]^{1/2}$.  After differentiating
w.r.t.~$s_0$ we find
\begin{align}\label{eqn:integrand}
  J'_s(m^2) &= \int_0^1 dx \int \frac{d^4p}{(2\pi)^4} \frac{[{\cal
      F}_1(m^2,p^2)+{\cal F}_2(m^2,p^2){\cal G}_s(p_0 V^{1/4})] \exp(i
      L^s p_0)}{\Bigl(p_0 - i\sqrt{p_\perp^2+m^2}\Bigr)\Bigl(p_0 +
    i\sqrt{p_\perp^2+m^2}\Bigr)}
\end{align}
with
\begin{align}
  {\cal G}_s(p_0 V^{1/4}) 
  &=\frac{i p_0}{L^s} -\frac{i p_0 s_0^2 \tilde{L}_0^2}{(L^s)^3} - \frac{p_0^2 s_0^2\tilde{L}_0^2}{(L^s)^2}
\end{align}
and
\begin{align}
  p^2 = p_0^2 + p_\perp^2\,.
\end{align}
In Figure~\ref{fig:p0cplane} we sketch the structure of the integrand of
Eq.~\eqref{eqn:integrand} in the complex plane.  \FIGURE[!t]{
  \centering
  \includegraphics{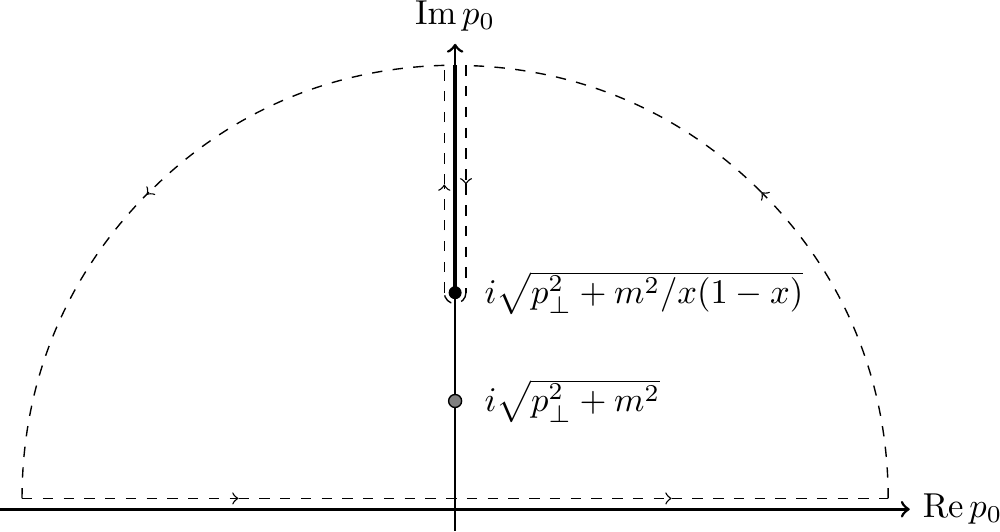}
  \caption{The complex plane of $p_0$.}
  \label{fig:p0cplane}
}
There are two poles at $p_0 = \pm i \sqrt{p_\perp^2+m^2}$ and a branch
cut due to the logarithms in ${\cal F}_1(m^2,p^2)$ and ${\cal F}_2(m^2,p^2)$.
We can close the integration contour in the upper half-plane and find
\begin{align}
  J'_s(m^2) = [J'_s(m^2)]_p + [J'_s(m^2)]_c\,,
\end{align}
where $[J'_s(m^2)]_p$ is the contribution of the pole and $[J'_s(m^2)]_c$ is the
contribution of the branch cut.  The contribution of the pole is given by
\begin{align}
  [J'_s(m^2)]_p &=\frac1{(2\pi)^2 V}\int_0^1 dx \int_0^\infty d\hat p_\perp
  \hat p^2_\perp \exp\Bigl(- l^s \sqrt{\hat p_\perp^2+m^2\sqrt V}\Bigr)\notag\\
&\quad\times\frac{{\cal F}_1(m^2,-m^2) \sqrt{V}+{\cal F}_2(m^2,-m^2){\cal
      G}_s\Bigl(i \sqrt{\hat p_\perp^2+m^2
      \sqrt{V}}\Bigr)\sqrt{V}}{\sqrt{\hat p_\perp^2+m^2 \sqrt{V}}}
\end{align}
with $l^s = L^s/V^{1/4}$ and $\hat p_\perp=p_\perp V^{1/4}$.  The contribution of the branch cut is given
by
\begin{align}
  [J'_s(m^2)]_c =\frac1{(2\pi)^4}\int_0^1 dx& \int d^3p_\perp \int_{i\sqrt{p_\perp^2+m^2/x(1-x)}}^{i\infty} dp_0 \exp(i L^s p_0) \notag\\
  &\times \frac{\Disc {\cal F}_1(m^2,p^2)+\Disc {\cal F}_2(m^2,p^2){\cal
      G}_s(p_0 V^{1/4})}{p^2+m^2}\,,
\end{align}
where
\begin{align}
  \Disc {\cal F}_1(m^2,p^2) &= \lim_{\varepsilon\to 0} \left[ {\cal
      F}_1(p_\perp^2+(p_0+\varepsilon)^2) - {\cal
      F}_1(p_\perp^2+(p_0-\varepsilon)^2) \right] \notag \\
  &= 2\pi i[m^2+x(1-x)p^2]\,,\notag\\
  \Disc {\cal F}_2(m^2,p^2) &= \lim_{\varepsilon\to 0} \left[ {\cal
      F}_2(p_\perp^2+(p_0+\varepsilon)^2) - {\cal
      F}_2(p_\perp^2+(p_0-\varepsilon)^2) \right] \notag \\
  &=4 \pi i x^2 \,.
\end{align}
Therefore
\begin{align}
  [J'_s(m^2)]_c = \frac{2}{(2\pi)^2 V}&\int_0^1 dx \int_0^\infty d\hat p_\perp
  \hat p_\perp^2
  \int_{\sqrt{\hat p_\perp^2+m^2 \sqrt{V}/x(1-x)}}^{\infty} dy   \exp(- l^s y)\notag\\
  &\times\frac{m^2 \sqrt{V}+x(1-x)(\hat p_\perp^2-y^2)+2 x^2 {\cal G}_s(i
    y)\sqrt{V}}{y^2-\hat p_\perp^2-m^2 \sqrt{V}}
\end{align}
with $p_0=iy$ and thus $dp_0 = i dy$.  In Sec.~\ref{sec:optnf2} we
calculate $[P_6^{1C}(m^2)]_\text{finite}$ numerically at scale $\mu=V^{-1/4}$,
i.e., we replace ${\cal F}_1(m^2,p^2)$ and ${\cal F}_2(m^2,p^2)$ by
\begin{align}
  {\cal F}_1(m^2,p^2) &= [m^2+x(1-x)p^2]\log\bigl[m^2\sqrt{V}+x(1-x) p^2\sqrt{V}\bigr]\,,\notag\\
  {\cal F}_2(m^2,p^2) &= 2 x^2 \bigl(1+\log[m^2\sqrt{V}+x(1-x)p^2\sqrt{V}]\bigr)\,.
\end{align}

\subsubsection{The term $P_6^{1D}(m^2)$}
The term $P_6^{1D}(m^2)$ is equal to the term $P_6^{1C}(m^2)$.  This can be seen
by shifting the integration variables $p_\mu \to p_\mu - k_\mu$ and
using the invariance of the integral under $k_\mu \to -k_\mu$.

\subsubsection{The term $P_6^{1E}(m^2)$}\label{app:p61e}
The term $P_6^{1E}(m^2)$ is finite and can be calculated numerically.  We
rewrite $P_6^{1E}(m^2)$ analogously to $P_6^{1B2}(m^2)$ as
\begin{align}
  P_6^{1E}(m^2) &=-\frac1{(8\pi)^2} \sum_{r \ne 0, s \ne 0,
    r \ne s} \int_0^\infty dx dy dz \: \frac{1}{(x y+x z+y z)^{3}} \notag\\
&\quad\times
  \biggl[2 (x + y) - \frac{({\tilde{L}_0}^2/\sqrt{V}) [-2 s_0 y + 2 r_0 (x +
      y)]^2\pi}{y z + x y + x z} \biggr]\notag \\&\quad\times \exp\biggl[ -
    \sum_j\frac{({\tilde{L}_j}^2/\sqrt{V}) [-2 r_j s_j y + r_j^2 (x + y) + s_j^2 (y +
      z)]\pi}{y z + x y + x z}\biggr] \notag\\&\quad
\times\exp\biggl[-(x+y+z) \frac{m^2 \sqrt{V}}{4 \pi}
  \biggr]\,.
\end{align}
In the following we discuss how to efficiently calculate the infinite
sums over $s_0,\ldots,s_3,r_0,\ldots,r_3$.  We first define
\begin{align}
  g &= \sum_{r \ne 0, s \ne 0, r \ne s} \prod_{j=0}^3\exp(-a_j r_j^2-b_j s_j^2+c_j r_j s_j+d_j r_j+ e_j s_j) \notag\\
  &=\sum_{s \ne 0} \biggl[\sum_r\prod_{j}\exp(-a_j r_j^2-b_j s_j^2+c_j r_j s_j
  + d_j r_j+e_j s_j) - \prod_{j}\exp(-b_j s_j^2+e_j s_j) \notag\\&\qquad\quad\qquad
  - \prod_{j}\exp[-(a_j+b_j-c_j)s_j^2+(d_j+e_j)s_j]\biggr] \notag\\
  &=2+\sum_{r,s}\prod_{j} \exp(-a_j r_j^2-b_j s_j^2+c_j r_j s_j + d_j r_j+e_j
  s_j) - \sum_s\prod_{j}\exp(-b_j s_j^2+e_j s_j) \notag\\&\quad -\sum_r
  \prod_{j}\exp(-a_j r_j^2 + d_j r_j) -
  \sum_s\prod_{j}\exp[-(a_j+b_j-c_j)s_j^2+(d_j+e_j)s_j]\,,
\end{align}
where $a_j, b_j, c_j, d_j, e_j \in \mathbbm{C}$ with $\re a_j, \re b_j
>0$ and $j=0,\ldots,3$.  In this way we can write
\begin{align}
  g 
&= \prod_{j=0}^3 \bar g(a_j,b_j,c_j,d_j,e_j) - \prod_{j=0}^3\nu(b_j,e_j)-\prod_{j=0}^3\nu(a_j,d_j) 
-\prod_{j=0}^3\nu(a_j+b_j-c_j,d_j+e_j)+2\,,
\end{align}
where
\begin{align}\label{eqn:nudef}
  \nu(a,b) &=\sum_{n} \exp(-a n^2 + b n)\,,\notag\\
  \bar g(a,b,c,d,e) &= \sum_{n,m} \exp(-a n^2-b m^2 + c m n + d n + e m) \notag\\&= \sum_m\nu(a,c m+d) \exp(-b m^2+e m)\,.
\end{align}
Since $\nu$ is a Jacobi theta function it transforms covariantly under inversion of $a$,
\begin{align}\label{eqn:nuinv}
  \nu(a,b)&=\sqrt{\frac{\pi}{a}}\exp(b^2/4a)\nu(\pi^2/a,i \pi (b/a))\,,
\end{align}
which is readily shown using Poisson's summation formula.  For $a \in
\mathbbm{R}$ with $a < \pi$ we can therefore use Eq.~\eqref{eqn:nuinv}
to achieve a swift convergence of the sum over $n$ in $\nu(a,b)$ in
Eq.~\eqref{eqn:nudef}.  In fact in the least favorable case of $a =
\pi$ we need only to sum all $n$ with $\abs{n-n_0}\le 4$, where $n=n_0
\in \mathbbm{Z}$ maximizes $-a n^2+b n$, to achieve a precision of
$-\log_{10}[\exp(-4^2 \pi)] \approx 22$ digits.  In this way we can
express $P_6^{1E}(m^2)$ as
\begin{align}
  P_6^{1E}(m^2) 
&=-\frac1{(8\pi)^2} \int_0^\infty dx dy dz \: R^3
  \left[2 (x + y) - R \tilde l_0^2 [-2 y \partial_{e_0} + 2 (x +
      y) \partial_{d_0}]^2\pi \right]\notag \\&\quad\times
  g(e_0,d_0)\Big\vert_{e_0 = d_0 = 0} \exp\biggl[-(x+y+z) \frac{m^2 \sqrt{V}}{4 \pi}
  \biggr]
\end{align}
with
\begin{align}
R &= \frac1{y z + x y + x z}\,,&
\tilde l_j &= \tilde{L}_j/V^{1/4}\,,\notag\\
a_j &= R \tilde l_j^2 (x + y)\pi\,, &
b_j &= R \tilde l_j^2 (y + z)\pi\,,\notag\\
c_j &= R \tilde l_j^2 2\pi y\,, &
d_1 &= d_2 = d_3 = e_1 = e_2 = e_3 = 0\,.
\end{align}
The integral over $x$, $y$ and $z$ can be performed conveniently in
spherical coordinates, as in the case of $P_6^{1B2}(m^2)$.

\subsection{The complete diagram}\label{app:complete}
We combine all contributions to $P_6(m^2)$ and find that the complete
diagram at scale $\mu=V^{-1/4}$ is given by
\begin{align}
  (P_6)_\text{UV} &=\frac13 \lambda -\frac{10}3 \lambda P_1\,, \notag\\
  P_6^r(m^2) &=-\frac1{m^2\sqrt V}\tilde{L}_0\partial_{\tilde{L}_0}\beta_1 -\frac{\beta_1}{2m^2 \sqrt V}
  -\frac{\log(m^2 \sqrt V)}{(4\pi)^2}P_1 -\partial_{\tilde{L}_0}(\tilde{L}_0\beta_2) \notag\\  &\quad
 -\frac1{2(4\pi)^2}
  +2[P_6^{1C}(m^2)]_\text{finite} 
  -\frac{1}{(4\pi)^2}P_1 + P_6^{1B2}(m^2)
  + P_6^{1E}(m^2)\,,\notag\\
  P_6^r &= \lim_{m^2 \to 0} P_6^r(m^2)\,,\label{eq:p6}
\end{align}
where $P_6^r$ is the renormalized shape coefficient at scale
$\mu=V^{-1/4}$.  In Figs.~\ref{fig:p6a} and \ref{fig:p6b} we show the
result of a numerical calculation of $P_6^r(m^2)$ for different values
of $m^2\sqrt{V}$.

Note that the linear divergences as well as the logarithmic
divergences in $1/m^2\sqrt{V}$ cancel.  We perform a fit to a
polynomial of order four for different ranges of $m^2\sqrt{V}$: (i)
$m^2\sqrt{V}<0.75$, (ii) $m^2\sqrt{V}<1.5$, and (iii) $m^2\sqrt{V}<3$.
The result of the fits and the corresponding extrapolated values for
$m^2 = 0$ are given in Table~\ref{tab:fitsp6}. 

\newcolumntype{k}{D{.}{.}{2.8}}
\newcolumntype{l}{D{.}{.}{4}}

\TABLE[t]{
  \begin{tabular}{|c|kkk|k|}
    \hline
    & \multicolumn1l{\text{(i)}} & \multicolumn1l{\text{(ii)}} & \multicolumn1l{\text{(iii)}} & \multicolumn1{|c|}{\text{extrapolation}} \\\hline
    $(a_1)$ & -0.010591 & -0.010538 & -0.010501 & -0.0106(1) \\
    $(a_{3/2})$ & 0.009395 & 0.009429 & 0.009439 & 0.00940(4) \\
    $(a_2)$ & 0.024867 & 0.024892 & 0.024884 & 0.02487(2)\\
    $(a_3)$ &  0.031944 & 0.031799 & 0.031778 & 0.0319(1) \\
    $(a_4)$ & -0.021362 & -0.021531 & -0.021720 & -0.0214(3) \\
    $(b_{3/2})$ & -0.015129 & -0.015092 & -0.015050 & -0.01513(3) \\
    $(b_2)$ & -0.015765 & -0.015714 & -0.015669 & -0.01577(7) \\
    $(b_3)$ & -0.008005 & -0.007968 & -0.007904  & -0.0080(1)\\
    $(b_4)$ &  0.011724 & 0.011748 & 0.011810 & 0.0117(1) \\\hline
  \end{tabular}
  \caption{Best fit of $P_6^r$ for regions (i), (ii), and (iii) and
    extrapolated values of $P_6^r$ for different geometries.}
  \label{tab:fitsp6}
}

\FIGURE[p]{
  \begin{tabular}{rr}
    \includegraphics{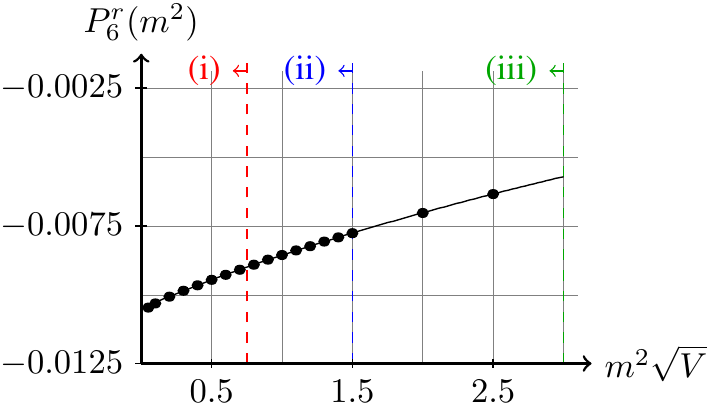} & \includegraphics{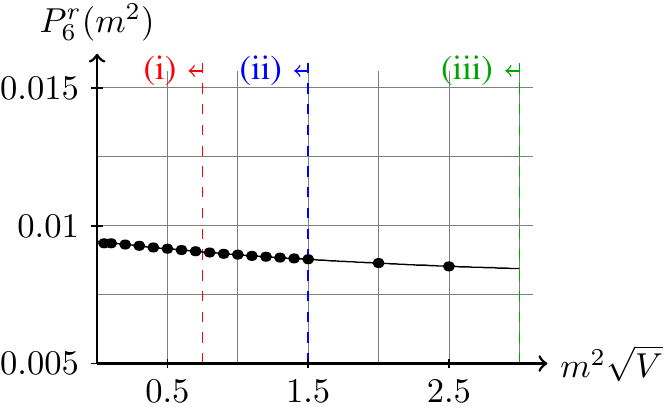} \\
    \multicolumn1c{$(a_{1})$} &       \multicolumn1c{$(a_{3/2})$} \\
    \includegraphics{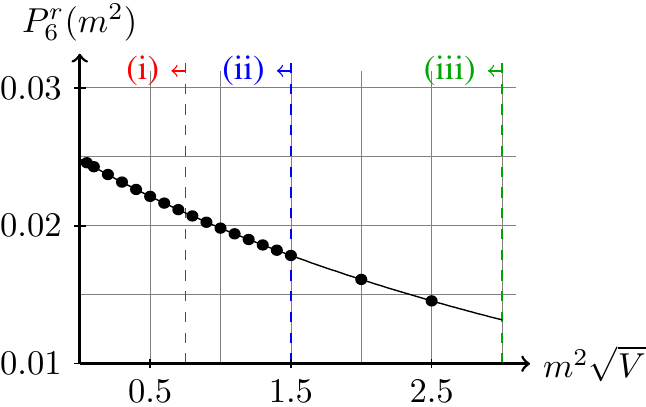} & \includegraphics{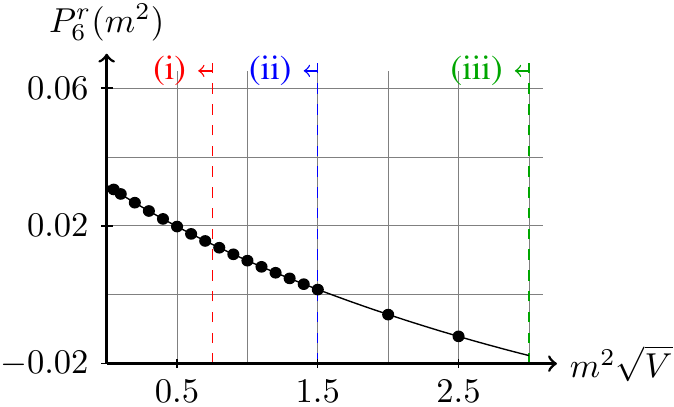} \\
    \multicolumn1c{$(a_{2})$} &      \multicolumn1c{$(a_{3})$} \\
    \includegraphics{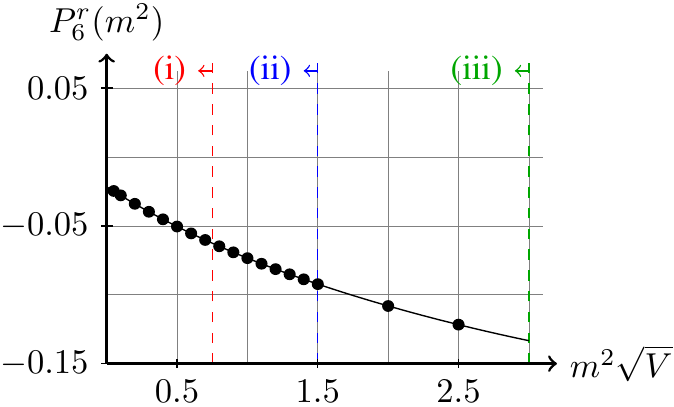} \\
    \multicolumn1c{$(a_{4})$}
  \end{tabular}
  \caption{Extrapolation of $P^r_6 = \lim_{m \to 0} P^r_6(m^2)$ for
    lattice geometries $(a_x)$, $x \in \{1,3/2,2,3,4\}$ at scale
    $\mu=V^{-1/4}$.  We fit a polynomial of order four including numerical
    data from three different ranges (i), (ii), and (iii).}
  \label{fig:p6a}
}

\clearpage

\FIGURE[t]{
  \begin{tabular}{rr}
    \includegraphics{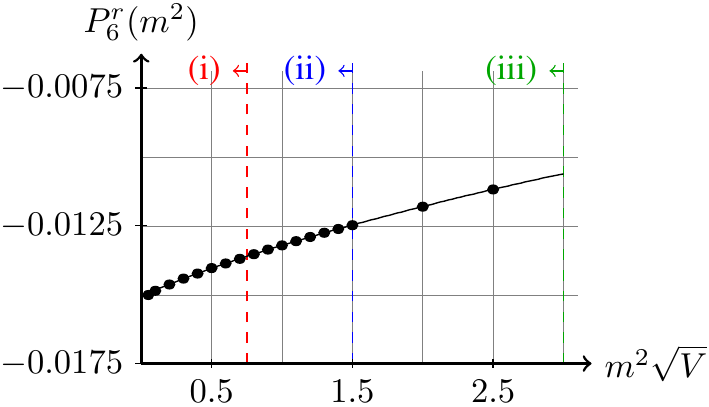} & \includegraphics{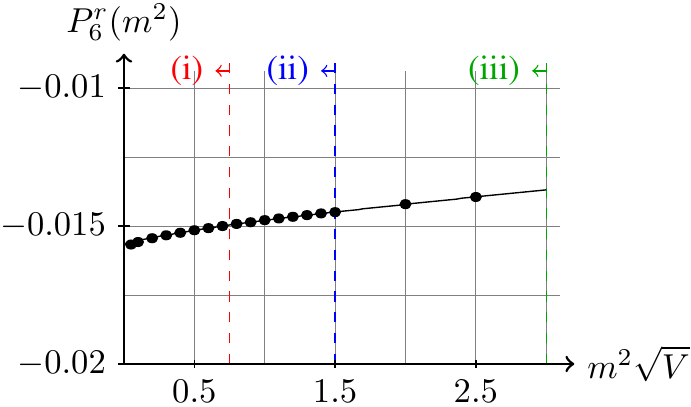}\\
    \multicolumn1c{$(b_{3/2})$} &    \multicolumn1c{$(b_{2})$} \\
    \includegraphics{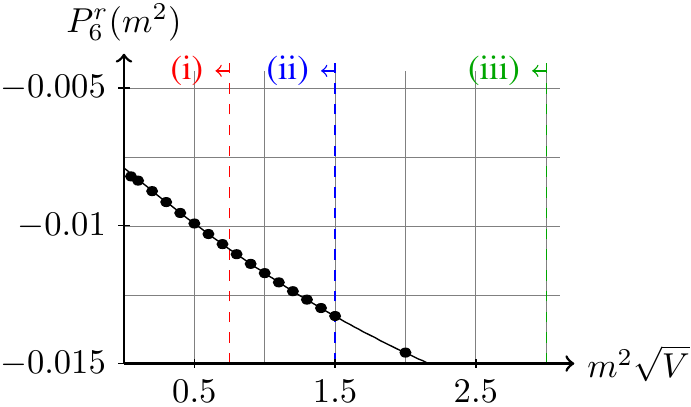} & \includegraphics{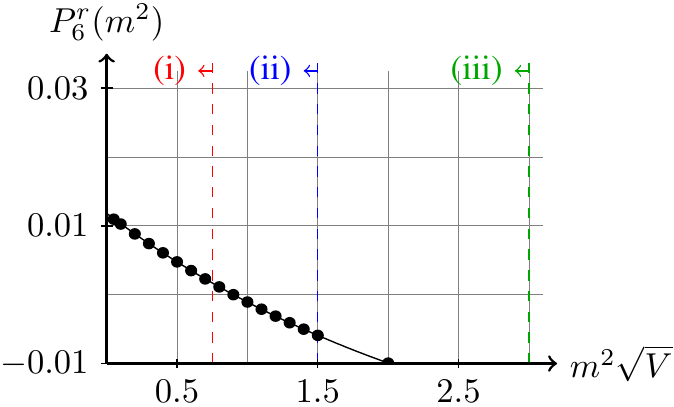}\\
    \multicolumn1c{$(b_{3})$} &    \multicolumn1c{$(b_{4})$}
  \end{tabular}
  \caption{Extrapolation of $P^r_6 = \lim_{m \to 0} P^r_6(m^2)$ for
    lattice geometries $(b_x)$, $x \in \{3/2,2,3,4\}$ at scale
    $\mu=V^{-1/4}$.  We fit a polynomial of order four including numerical
    data from three different ranges (i), (ii), and (iii).}
  \label{fig:p6b}
}


\bibliographystyle{JHEP}
\bibliography{nnlo} 

\end{document}